\documentclass[twocolumn,showpacs,preprintnumbers,amsmath,amssymb]{jpsj2}

\usepackage{graphicx}
\usepackage{dcolumn}
\usepackage{bm}
\usepackage{wrapfig}
\usepackage{color}
\usepackage{cases}

\title{
Potential antiferromagnetic fluctuations in hole-doped iron-pnictide superconductor Ba$_{1-x}$K$_{x}$Fe$_{2}$As$_{2}$
studied by $^{75}$As nuclear magnetic resonance 
}

\author{
Masanori Hirano$^{1}$~\thanks{mas.vt@graduate.chiba-u.jp}, Yuji Yamada$^{1}$, Taku Saito$^{1}$, Ryo Nagashima$^{1}$, Takehisa Konishi$^{2}$, \\
Tatsuya Toriyama$^{1}$, Yukinori Ohta$^{1,5}$, Hideto Fukazawa$^{1,5}$, Yoh Kohori$^{1,5}$, Yuji Furukawa$^{3}$, \\
Kunihiro Kihou$^{4,5}$, Chul-Ho Lee$^{4,5}$, Akira Iyo$^{4,5}$, and Hiroshi Eisaki$^{4,5}$
}
\inst{
$^{1}$Department of Physics, Chiba University, Chiba 263-8522, Japan\\
$^{2}$Department of Chemistry, Chiba University, Chiba 263-8522, Japan\\
$^{3}$Ames Laboratory and Department of Physics and Astronomy, Iowa State University, Ames, IA 50011, USA\\
$^{4}$National Institute of Advanced Industrial Science and Technology, Tsukuba 305-8568, Japan\\
$^{5}$JST, Transformative Research-Project on Iron Pnictides (TRIP), Chiyoda, Tokyo 102-0075, Japan\\
}
\date{\today}
\abst{
We have performed $^{75}$As nuclear magnetic resonance (NMR) and 
nuclear quadrupole resonance (NQR) on single crystalline 
Ba$_{1-x}$K$_{x}$Fe$_{2}$As$_{2}$ for $x$ = 0.27-1. 
$^{75}$As nuclear quadruple resonance frequency ($\nu_{\rm Q}$) increases linearly with increasing $x$. 
The Knight shift $K$ in normal state shows Pauli paramagnetic behavior with slight temperature $T$ dependence. 
The value of $K$ increases gradually with increasing $x$. 
By contrast, nuclear spin-lattice relaxation rate $1/T_{1}$ in normal state has a strong $T$ dependence, 
which indicates existence of large antiferomagnetic (AF) spin fluctuations for all $x$. 
The $T$ dependence of $1/T_{1}$ shows a gap-like behavior below approximately 100~K for $0.6 < x < 0.9$. 
These behaviors are well explained by the change of band structure with expansion of hole Fermi surfaces 
and shrink and disappearance of electron Fermi surfaces at Brillouin zone (BZ) with increasing $x$. 
The anisotropy of $1/T_{1}$, represented by a ratio of $1/T_{1ab}$ to $1/T_{1c}$, is always larger than 1 for all $x$, 
which indicates that the stripe-type AF fluctuations is dominant in this system. 
The $K$ in superconducting (SC) state decreases, 
which corresponds to appearance of spin-singlet superconductivity. 
The $T$ dependence of $1/T_{1}$ in SC state indicates multiple-SC-gap feature. 
A simple two gap model analysis shows that the larger superconducting gap gradually decreases 
with increasing $x$ from 0.27 to 1 and smaller gap decreases rapidly and 
nearly vanishes for $x > 0.6$ where the electron pockets in BZ disappear. 
}
\kword{iron pnictide superconductor, nuclear magnetic resonance, nuclear quadrupole resonance, spin fluctuations, superconducting gap}
\begin{document}
\maketitle

\section{Introduction}

Soon after the discovery of LaFeAsO$_{1-x}$F$_{x}$ with superconducting (SC) transition temperature 
$T_{\rm c}$ = 26~K by Kamihara {\it et al}. in 2008~\cite{Kam1}, 
iron based superconductors have been studied extensively all over the world. 
Highest superconducting transition temperature $T_{\rm c}$ of 56~K was 
soon observed in {\it R}FeAsO~\cite{Ren1,Miy1} (``1111'', {\it R} is a rare-earth element), 
which is a high $T_{\rm c}$ discovered next to the value in cuprate system. 
Many iron based superconductors were observed one after another in {\it A}FeAs 
(``111'' {\it A} is an alkaline element), Fe$_{1+\delta}$Se (``11''), 
and {\it AE}Fe$_{2}$As$_{2}$ (``122'', {\it AE} is an alkaline-earth element). 
Among them, {\it AE}Fe$_{2}$As$_{2}$ ({\it AE} = Ba, Ca, Sr) occupies a singular position, 
since high quality and large single crystals were grown. 
A parent compound BaFe$_{2}$As$_{2}$ has ThCr$_{2}$Si$_{2}$ type crystal structure and 
has an antiferromagnetic (AF) and tetragonal to orthorhombic crystal structure phase transition at 140~K. 
The orthorhombic and AF phase was suppressed and superconductivity appeared 
in hole doped Ba$_{1-x}$K$_{x}$Fe$_{2}$As$_{2}$~\cite{Rot1}, 
electron doped Ba(Fe$_{1-x}$Co$_{x}$)$_{2}$As$_{2}$~\cite{Sef1}, 
and BaFe$_{2}$(As$_{1-x}$P$_{x}$)$_{2}$~\cite{Jia1,Kas1} 
in which isovalent P substitution for As acts as chemical pressure. 
Indeed, the direct application of pressure for BaFe$_{2}$As$_{2}$ also induces 
superconductivity~\cite{Ali1,Yam1}.

As for SC gap structure, thermal conductivity, magnetic penetration-depth measurements 
in Ba(Fe$_{1-x}$Co$_{x}$)$_{2}$As$_{2}$ have shown existence of nodes 
in underdoped and overdoped region~\cite{Rei1}.
For BaFe$_{2}$(As$_{1-x}$P$_{x}$)$_{2}$, angle resolved thermal conductivity suggested the closed nodal loops 
located at the flat parts of the electron Fermi surface~\cite{Yam10}, 
while three dimensional nodal structure on hole Fermi surface was proposed 
by theoretical calculation using fluctuation exchange (FLEX) approximation~\cite{Suz1}.

In Ba$_{1-x}$K$_{x}$Fe$_{2}$As$_{2}$ (BKFA), many experiments revealed 
appearance of multiple full SC gap around $x =$ 0.4 
where $T_{\rm c}$ has a maximum value of 38~K~\cite{Has2,Kha1,Din1}. 
Sign changing $s_{\pm}$-wave which is mediated by spin fluctuations 
well explains a lot of experimental results~\cite{Maz1,Ike1,Nag1,Suz2}.
$T$ dependence of spin-lattice relaxation rate 1/$T_{1}$ 
can also be explained by $s_{\pm}$-wave~\cite{Yam20,Mata1}.
However, it is pointed out theoretically that $s_{\pm}$-wave is very fragile 
to nonmagnetic impurity~\cite{Ona1}, while iron-pnictide superconductors are experimentally 
robust against nonmagnetic impurity.
No sign changing $s_{++}$-wave which is mediated by orbital fluctuations and 
is robust against nonmagnetic impurity is another possible candidate 
for the Cooper pairing in this system~\cite{San1,Yan1,Ona2,Kon10}.
Hence, the SC pairing symmetry in optimum region is still an open question. 
However, there is consensus that the SC gap structure itself is fully gapped one.

By contrast, NQR and specific heat, thermal conductivity, and magnetic penetration-depth measurements 
for $x =$ 1 (KFe$_{2}$As$_{2}$) revealed appearance of nodal SC gap~\cite{Fuk1,Has1,Don1}. 
Small angle neutron scattering (SANS) experiment has pointed out that stable isotropic hexagonal vortex lattice 
exists when magnetic fields applies parallel to crystal $c$ axis, 
which supports nodes in horizontal direction~\cite{Kaw1}. 
Recent muon spin relaxation measurements of KFe$_{2}$As$_{2}$ also supports the SANS results~\cite{KOh1}. 
However, recent angle resolved photoemission spectroscopy (ARPES) and specific heat measurements 
indicate nodes in vertical direction~\cite{Oka1,Aok1}. 
The horizontal nodal structure with $s_{\pm}$-wave~\cite{Suz2} and 
nodal-line SC gap structure with $d$-wave~\cite{Tho1} are theoretically proposed to this end compound. 
It is noteworthy that SC gap structures in BKFA are different 
in optimally doped $x =$ 0.4 and in heavily overdoped $x =$ 1.

Important feature of BKFA is a wide range realization of SC phase for 0.2 $< x \leq$ 1~\cite{Rot1}. 
In BKFA, the band structure changes associated with hole doping. 
For $x =$ 0.4, cylindrical hole Fermi surfaces exist around $\Gamma$ point and 
cylindrical electron surfaces around X points of the Brillouin zone (BZ). 
Good nesting condition between disconnected hole and 
electron Fermi surfaces enhances AF spin fluctuations and 
is favorable for superconductivity mediated by AF spin fluctuations. 
Within this picture, $s_{\pm}$-wave superconductivity is expected~\cite{Maz1,Ike1,Nag1,Suz2}. 
Interestingly, the inelastic neutron scattering revealed that 
there is incommensurate spin fluctuation ${\bm Q}$ = ((1$\pm 2\delta$)$\pi$,(1$\pm 2\delta$)$\pi$,0) 
with $\delta$ = 0.16 for $x =$ 1 at excitation energies above 3~eV~\cite{Lee1} 
where electron Fermi surfaces completely disappear and apparent good nesting condition between inter bands at Fermi level does not exit. 
On the other hand, there is another theoretical proposal that SC is mediated by orbital fluctuations,
which leads to nearly orbital independent SC gap $s_{++}$~\cite{San1,Yan1,Ona2,Kon10}. 
Hence, there remains much work to clarify the SC gap symmetry in iron pnictide superconductors. 
Band structure of BKFA changes associated with hole doping. 
The hole Fermi surfaces expand with increasing $x$, 
whereas electron Fermi surfaces shrink gradually and disappear for $x >$ 0.6~\cite{Nak1,Sat1,Ter1}. 
In BKFA, $T_{\rm c}$ changes smoothly with increasing $x$. 
It is very important to clarify the change of SC gap structures in BKFA 
with interpolating optimally doped and overdoped regions. 
The relation between AF spin fluctuations and $T_{\rm c}$ should be also studied. 
Recently, we succeeded in synthesizing high quality large single crystals of BKFA for $0.31 \leq x \leq 1$, 
which provides us a unique opportunity to solve above problems.

NMR and/or NQR is suitable for study of static and dynamical magnetic properties, 
and provides valuable information on SC gap symmetry and the gap structures 
through Knight shift ($K$) and spin lattice relaxation rate ($1/T_{1}$) measurements. 
In this paper, we report a $^{75}$As-NMR/NQR study of BKFA.

\section{Experimental}

Single crystals of BKFA ($x$ = 0.27, 0.31, 0.39, 0.58, 0.64, 0.69, 0.94, 1) were grown by the BaAs and KAs self-flux method. 
X-ray diffraction showed that the crystals had a tetragonal ThCr$_{2}$Si$_{2}$ type structure with no impurity phase. 
The detailed procedure of single crystal growth is basically the same described in ref.~\ref{Kihou} 
which reports the single crystal growth of an end member of the system, KFe$_{2}$As$_{2}$. 
The ratio of Ba and K was determined by the energy dispersive X-ray spectroscopy. 
Experimental error of the evaluated values was within 5\%. 
The $c$-axis parameter of the single crystals was determined by X-ray diffraction analysis and 
the relation between the composition $x$ and lattice parameter $c$ is consistent with the former reported relation 
between the nominal composition $x$ and lattice parameter by Rotter {\it et al.}~\cite{Rot1}. 

We determined the $T_{\rm c}$ and superconducting volume fraction of the samples with a commercial superconducting-quantum-interference-device magnetometer. 
The $T_{\rm c}$'s of the samples are 38.5~K ($x$ = 0.27), 36.5~K ($x$ = 0.31), 38~K ($x$ = 0.39), 30.5~K ($x$ = 0.58), 26.5~K ($x$ = 0.64), 
20.5~K ($x$ = 0.69), 4.5~K ($x$ = 0.94), 3.5~K ($x$ = 1) and  
the volume fraction for all the $x$'s is approximately 100\% except for $x$ = 0.27. 
For $x$ = 0.27, the fraction is approximately 60\% because the crystal contains the phase-separated antiferromagnetically ordered phase~\cite{Fuk2} 
as we report the experimental evidence of the phase separation below. 
We also determined the $T_{\rm c}$ in the NMR-measurement magnetic field by the change of inductance of NMR detecting coil with changing $T$. 

The NMR/NQR experiment on the $^{75}$As nucleus ($I=3/2$, $\gamma /2\pi = 7.292$~MHz/T) was carried out 
using phase-coherent pulsed NMR/NQR spectrometers and a superconducting magnet between approximately 3 and 7~T. 
The measurement was performed using a $^{4}$He cryostat.  
The NMR spectra were measured by sweeping the applied fields at a constant resonance frequency. 
Magnetic field was applied parallel to the crystal $ab$ plane and the $c$ axis. 
Field alignment was performed with the eye. 
The origin of the Knight shift $K=0$ of the $^{75}$As nucleus was determined by the $^{75}$As NMR measurement of GaAs~\cite{Bas1}. 
The NQR spectra were measured by sweeping the frequency in zero magnetic field. 
$T_{1}$ was measured by a saturation recovery method at center of As-spectrum. 
We obtained $T_{1}$ at a fixed frequency of 37.15 or 43.75~MHz with an external field of 5.06-5.09 or 5.9-5.96~T in the $T$ range of 2-300~K, respectively.

\section{Results and discussion}

\subsection{NQR frequency}

\begin{figure}
\includegraphics[width=8cm]{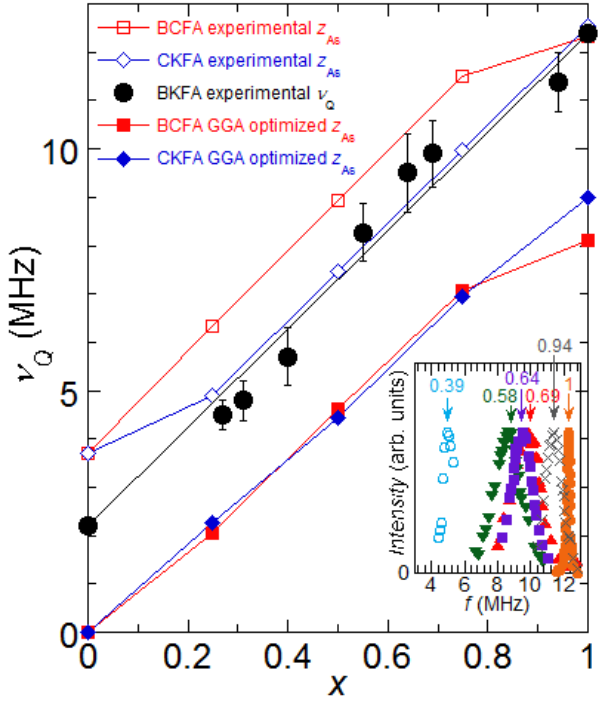}
\caption{
(color online) Nuclear quadrupole resonance frequency $\nu_{\rm Q}$ against concentration $x$. 
BCFA, CKFA, and BKFA are abbreviation for Ba$_{1-x}$Cs$_{x}$Fe$_{2}$As$_{2}$, Ca$_{1-x}$K$_{x}$Fe$_{2}$As$_{2}$, 
and Ba$_{1-x}$K$_{x}$Fe$_{2}$As$_{2}$, respectively. 
The experimental data of $\nu_{\rm Q}$ for $x=0$ were quoted from ref.~\ref{Kitag}. 
Closed circles are experimental results of $\nu_{\rm Q}$ obtained from $^{75}$As NQR measurements as shown in the inset. 
Open squares and diamonds are the results of band-structure calculation using the experimentally obtained $z$ parameter of As in ThCr$_{2}$Si$_{2}$ tetragonal structure.
$z$ parameter is $z$-coordinate of As site in the unit cell.
Closed squares and diamonds are the calculated results by optimizing the $z$ parameter with GGA. 
The NQR spectra in the inset were obtained at 40 ($x$ = 0.39 and 0.58), 35 ($x = 0.64$), 30 ($x = 0.69$), 10 ($x = 0.94$), and 4.2~K ($x = 1$). 
}
\label{f1}
\end{figure}

In Fig.~\ref{f1}, we show the NQR frequency $\nu_{\rm Q}$ against substitution concentration $x$.
The experimentally obtained $\nu_{\rm Q}$ was evaluated from the $^{75}$As NQR spectra of BKFA as shown in the inset of Fig~\ref{f1}. 
Bars for each data point correspond to the full width at half maximum ($FWHM$) of the spectra.
It is expected that the spectral broadening was brought out because of the random distribution of Ba$^{2+}$ and K$^{+}$ in the crystal. 
Here, $z_{\rm As}$ parameter indicates $z$-coordinate of As site in the unit cell. 

The $\nu_{\rm Q}$ generally depends on temperature and these spectra were obtained at the different measurement temperatures. 
However, note that such temperature dependence of $\nu_{\rm Q}$ in each compound is within the bar in Fig.~\ref{f1}. 
Hence, it is clear that the $x$ dependence of the experimental $\nu_{\rm Q}$ is nearly linear. 
The principal axis of the electric field gradient is along the crystal $c$-axis 
since the As site has a local fourfold symmetry around the $c$-axis.
Hence, asymmetry parameter $\eta$ at As site is basically $\eta = 0$.
However, we will discuss the possibility of finite $\eta$ in the following subsection \S~\ref{NMRspec}.
The definition of $\nu_{\rm Q}$ at As site in these materials is as follows: \\
\begin{equation}
\nu_{\rm Q} = \frac{3e^{2}qQ}{2hI(2I-1)}, 
\end{equation}
where $h$, $eq$, $eQ$ represent the Planck constant, the electric field gradient (EFG), and the nuclear quadrupole moment, respectively. 
The main contribution of change in $\nu_{\rm Q}$ can be simply interpreted as a linear increase of effective ligand valency by substitution of K for Ba in addition to the change in lattice parameter.
These two changes directly bring out the change in the EFG. 

This tendency is also confirmed by the systematics obtained in the calculated electronic structures for BKFA. 
We performed the electronic structure calculation using WIEN2K code~\cite{Bla1}, where the full-potential linearized augmented-plane-wave method (FLAPW) 
with the generalized gradient approximation (GGA) for electron correlations is used.
We also used the virtual crystal approximation for alloying effect, 
where we made calculations for hypothetical materials Ba$_{1-x}$Cs$_{x}$Fe$_{2}$As$_{2}$ and Ca$_{1-x}$K$_{x}$Fe$_{2}$As$_{2}$ 
to see the electronic structure of Ba$_{1-x}$K$_{x}$Fe$_{2}$As$_{2}$.
We used experimental lattice parameters $a$ and $c$ reported in ref.~\ref{Rott}.
The obtained Fermi surfaces using the experimental $z$ parameter of As in the ThCr$_{2}$Si$_{2}$ tetragonal structure are 
consistent with Fermi surfaces reported by de Haas-van Alphen effect~\cite{Ter1} and ARPES~\cite{Din1,Sat1,Nak1}. 
The Fermi surfaces obtained using the optimized value of the $z$ parameter was not consistent for the end member of the system:
electron Fermi surfaces do not disappear at the corner X point of the BZ even for $x = 1$.
This tendency is already reported by Singh with the same procedure using local density approximation~\cite{Sin1}. 

In this system changes in the lattice parameters, Fe-Fe interatomic distance, and the position of the As ions upon doping are 
closely interconnected with the electronic structure. 
In all the calculated results, $\nu_{\rm Q}$ exhibits monotonic increase with $x$ and has a same order of the value compared with the experimental $\nu_{\rm Q}$.  
Of all the results, it is surprising that the experimental $\nu_{\rm Q}$ has a good agreement with the $\nu_{\rm Q}$ 
calculated for Ca$_{1-x}$K$_{x}$Fe$_{2}$As$_{2}$ in the virtual crystal approximation 
since such agreement can rarely be seen in other strongly electron-correlated compounds. 
Such good agreement between experiments and theory suggests that 
the electronic structure calculated assuming the experimental $z$ parameters of As can explain 
the experimental electronic state of BKFA fairly well.

\subsection{NMR spectra}
\label{NMRspec}

\begin{figure}
\includegraphics[width=8cm]{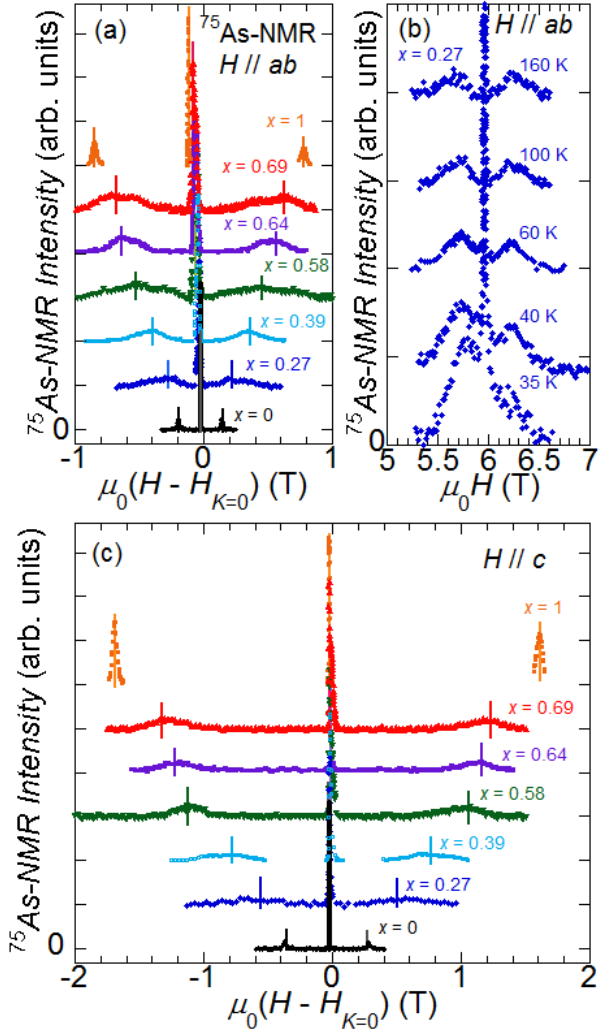}
\caption{
(color online) The NMR spectra of BKFA with magnetic field (a) parallel and (c) perpendicular to the $ab$ plane 
above $T_{\rm c}$ or $T_{\rm N}$. 
The data for $x=0$ was quoted from ref.~\ref{Kitag}. 
The measurement temperatures are 141, 100, 60, 50, 30, 25 and 4.2~K for $x$ = 0, 0.27, 0.39, 0.58, 0.64, 0.69 and 1, respectively. 
Solid lines denote the fitted parameter obtained from the analysis described in the text. 
(b) The temperature dependent NMR spectra of $x = 0.27$.
}
\label{f2}
\end{figure}

In Fig.~\ref{f2}(a) and (c), we show the NMR spectra of BKFA with magnetic field parallel and perpendicular to the $ab$ plane 
above $T_{\rm c}$ or $T_{\rm N}$. 
In Fig.~\ref{f2}(b), we show the temperature dependent NMR spectra of $x = 0.27$. 
We used the NMR data for $x=0$ from ref.~\ref{Kitag}. 
Because $^{75}$As nucleus has quantum number $I = 3/2$, the nuclear-spin Hamiltonian 
with an external magnetic field $H_{\rm ext}$ is given by 
\begin{equation}
 \mathcal{H} = -\gamma H_{\rm ext}I_{z} -\gamma \vec{H}_{\rm ext}\cdot\bar{K}\cdot\vec{I} 
  +\frac{\nu_{\rm Q}}{6}\left( 3I_{z'}^{2}-I(I+1)\right), 
\end{equation}
$$(\vec{H}_{\rm ext}\cdot\bar{K}\cdot\vec{I})/H_{\rm ext} = -(K_{ab}\sin\theta)(I_{x}\cos\theta-I_{z}\sin\theta)$$
\begin{equation}
 \qquad\qquad\qquad\qquad\quad\;\,+(K_{c}\cos\theta)(I_{x}\sin\theta+I_{z}\cos\theta),
\end{equation}
\begin{equation}
 I_{z'} = I_{x}\sin\theta+I_{z}\cos\theta,
\end{equation}
where $\bar{K}$ ($K_{ab}$ and $K_{c}$) and $\theta$ represent the Knight shift tensor (Knight shifts) 
and the angle between quantization axis and the crystal $c$ axis, respectively.
Here, we took the direction of the external magnetic field as principal axis of quantization axis 
and we also assumed the diagonalized $\bar{K}$ and the absence of in-plane anisotropy of $K_{ab}$. 
The detailed procedure to obtain $K_{ab}$ and $K_{c}$ from the NMR spectra is described elsewhere~\cite{Fuk3}.

The $FWHM$ of the center peak for $x=1$ is about 48 $\pm$ 2~kHz.
Hence, the analysis using the above expression works well for $x = 1$ and 
it turned out that the actual field misalignment was within approximately 5\% accuracy~\cite{Fuk3}.
But we used the following approximation for the other $x$'s since the NMR spectra are much broader than that of $x = 1$.
\begin{equation}
 \mathcal{H} \simeq -\gamma (1+K_{ab})H_{\rm ext}I_{z} +\frac{\nu_{\rm Q}}{6}\left( 3I_{x}^{2}-I(I+1)\right)\; (H//ab), 
\end{equation}
\begin{equation}
 \mathcal{H} \simeq -\gamma (1+K_{c})H_{\rm ext}I_{z} +\frac{\nu_{\rm Q}}{6}\left( 3I_{z}^{2}-I(I+1)\right)\; (H//c). 
\end{equation}
Even with the approximation, we cannot determine $K_{ab}$ for $x=0.58$, 0.64, and 0.69 with sufficient accuracy 
because all the center and satellite lines are broad and the position of the center line is affected by both the $K_{ab}$ and $\nu_{\rm Q}$: 
the $FWHM$ of the center peak for these composition is approximately 4-8 times broader than that for $x=1$. 
Therefore, the solid lines in Fig.~\ref{f2}(a) for $x=0.58$, 0.64, and 0.69 are just reference containing large fitting error. 
We can determine the $K_{c}$ for $x=0.58$, 0.64, and 0.69 with reasonable fitting error 
since the position of the center line is affected only by the $K_{c}$ and the line width of the center line was rather sharper: 
the $FWHM$ of the center peak for these composition is approximately 2-3 times broader than that for $x=1$. 
The satellite lines of $x$ = 0.58, 0.64, and 0.69 seem to have small different structures.
It probably arises from appearance of local asymmetries by K doping.
Because $\eta$ does not affect the central transition,
we assume $\eta = 0$ for the following discussion.

The NMR spectra of $x = 0.27$ are symmetric at 100 and 160~K. 
However, the overall feature of the spectra slightly changes below about 60~K. 
In addition to the symmetric center and satellite lines, broad peak appears: 
the position of this peak is just below the center position. 
Below $T_{\rm c}$ this feature becomes obvious. 
This is because the paramagnetic region and the antiferromagnetically ordered region are phase-separated in this compound. 
This phenomenon was previously reported for polycrystalline samples~\cite{Fuk2}.
However, recent  $\mu$SR study of underdoped BKFA shows a microscopic coexistence of SC and AF phases without phase separation~\cite{Wie1}.

\subsection{Knight shifts}

\begin{figure}
\includegraphics[width=8cm]{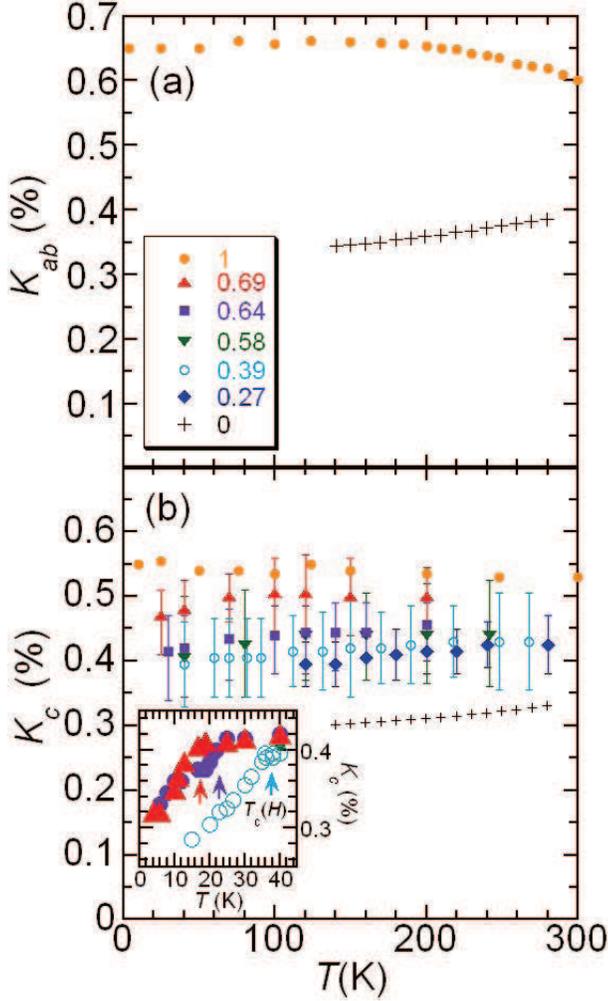}
\caption{
(color online) The temperature dependence of Knight shift of BKFA 
with magnetic field (a) parallel and (b) perpendicular to the $ab$ plane. 
The data for $x=0$ was quoted from ref.~\ref{Kitag}. 
In the inset, the temperature dependence of $K_{c}$ below 30~K is shown for $x$ = 0.39, 0.64, 0.69. 
}
\label{f3}
\end{figure}

In Fig.~\ref{f3}(a) and (b), we show the temperature dependence of Knight shift $K(T)$ of BKFA 
with magnetic field parallel and perpendicular to the $ab$ plane. 
For all the data, the $K(T)$ exhibits weak temperature dependence. 
This tendency is similar to the isovalent substitution system BaFe$_{2}$As$_{2-x}$P$_{x}$~\cite{Nakai1}. 
Although we could not obtain the $K_{ab}$ for $x=0.58$, 0.64, and 0.69, 
systematic development of the $K_{c}$ indicates that the $K_{ab}$ for $x=0.58$, 0.64, and 0.69 exists between $x= 0.27$ and 1
and that it also exhibits weak temperature dependence. 
With increasing $x$, $K(T)$ at same temperature basically increases. 
This is attributable to the increase of density of states at the Fermi level with hole doping. 
Thus, static magnetic property is nearly temperature independent and shows the slight $x$ dependence: 
spin fluctuation at around $\vec{q} = \vec{0}$ is not significant in this system. 

The hyperfine coupling constant $A_{i}$ parallel or perpendicular to the $ab$-plane for $x=1$ was 
evaluated from the Knight shift at the As site parallel or perpendicular to the $ab$-plane 
and the magnetic susceptibility parallel or perpendicular to the $ab$-plane by assuming the following formula, 
\begin{eqnarray}
K_{i}(T) = \frac{A_{i}^{d}}{N_{\rm A}\mu_{\rm B}}\chi_{i}^{d}(T) + \frac{A_{i0}}{N_{\rm A}\mu_{\rm B}}\chi_{i0}, \\ 
\chi_{i}(T) = \chi_{i}^{d}(T) + \chi_{i0},\quad i=ab\; {\rm or}\; c.   
\end{eqnarray}
Here, $N_{\rm A}$ represents Avogadro's number. 
$\chi_{i}^{d}$ represents $T$ dependent spin susceptibility originating from 3d electron of Fe.
$\chi_{i0}$ represents $T$ independent terms arising from orbital contribution.
The evaluated hyperfine coupling constants $A_{ab}^{d}$ and $A_{c}^{d}$ originating from 
the transfer from 3$d$ conduction electron to $^{75}$As nuclear spin of Fe are +26(5) and +13(4)~kOe/$\mu_{\rm B}$, respectively.
These values are quite similar to those obtained for $x=0$~\cite{Kit1}. 
This indicates that the hyperfine coupling constant nearly remains constant with hole doping in this system. 

In the inset of Fig.~\ref{f3}(b), we show the temperature dependence of $K_{c}$ below 30~K for $x$ = 0.39, 0.64, 0.69. 
The $K_{c}$ exhibits clear decrease below $T_{\rm c}(H)$ for both the samples. 
Since the Knight shifts correspond to the spin susceptibility,
this result indicates spin-singlet superconductivity in BKFA for $x$ = 0.39, 0.64, 0.69. 
Combined with the results for $x$ = 1~\cite{Fuk3}, even in the heavily hole-doped region, the spin part of superconducting pairing symmetry is singlet 
as realized in other iron-pnictide superconductors~\cite{Mata1,Nin1,Mata2,Nak1} which are close to parent compensated metals.
Note that we could not find any anomaly of $K_{\rm c}$ associated with multiple SC gap structures.

\subsection{Spin lattice relaxation rate $1/T_{1}$ in normal state}

\begin{figure}
\includegraphics[width=8cm]{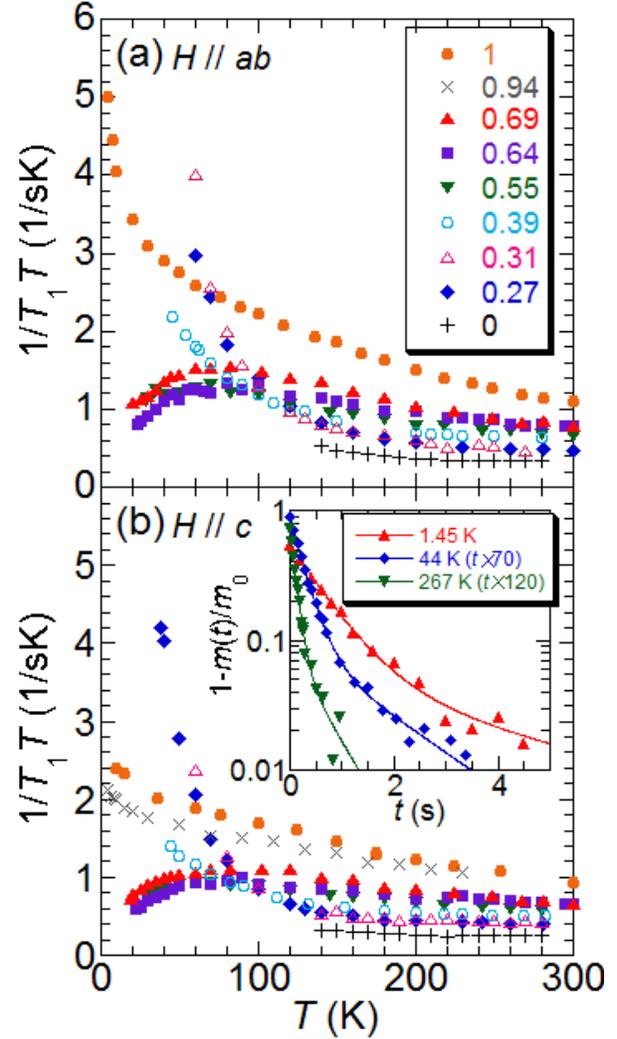}
\caption{
(color online) Temperature dependence of spin-lattice relaxation rate inversely multiplied with temperature 
$1/T_{1}T$ of $^{75}$As of BKFA
with magnetic field (a) parallel and (b) perpendicular to the $ab$ plane. 
The data for $x=0$ was quoted from ref.~\ref{Kitag}. 
The inset shows the typical recovery curve for $x = 0.69$ obtained at 1.45, 44, 267~K. 
}
\label{f4}
\end{figure}

In Fig.~\ref{f4}, we show the temperature dependence of spin-lattice relaxation rate divided by temperature 
$1/T_{1}T$ of $^{75}$As of BKFA
in the normal state with magnetic field (a) parallel and (b) perpendicular to the $ab$ plane. 
The nuclear magnetization recovery curve was fitted by the following double-exponential function 
as expected for the center line of the spectrum of the nuclear spin $I=3/2$ of the $^{75}$As nucleus~\cite{Sim1}, 
\begin{equation}
1-\frac{m(t)}{m_{0}} = 0.1\exp\left( -\frac{t}{T_{1}}\right)  + 0.9\exp\left( -\frac{6t}{T_{1}}\right),
\end{equation}
where $m(t)$ and $m_{0}$ are nuclear magnetizations after a time $t$ and enough time from the NMR saturation pulse. 
In the inset of Fig.~\ref{f4}, we show the typical recovery curve for $x$ = 0.69 obtained at 1.45, 44, 267~K. 
Clearly, the data are well fitted to the above ideal curve with a single $T_{1}$ component.  
We tried to use this formula for $x$ = 0.27 below about 100~K under magnetic field parallel to the $ab$ plane, 
but could not obtain good fitting results. 
This is due to more than two $T_{1}$ components below this temperature 
because the center line shown in Fig.~\ref{f2}(b) consists of the paramagnetic and AF states. 
For this composition parallel to the $c$ axis we could obtain $T_{1}$ down to $T_{\rm c}$ 
because the center line consists of only paramagnetic state. 

For all the composition, the $1/T_{1}T$ monotonically increases down to approximately 100~K. 
However, below this temperature there is clear difference. 
While the $1/T_{1}T$ continuously increases down to $T_{\rm c}$ for $x$ = 0.27, 0.31, 0.39, and 1, 
it saturates at around 60-80~K, and decreases below this temperature for $x$ = 0.58, 0.64, 0.69. 
This gap-like behavior for $x$ = 0.58, 0.64, 0.69 is 
remembrance of that in electron-doped system Ba(Fe$_{1-x}$Co$_{x}$)$_{2}$As$_{2}$~\cite{Nin1}. 
In the system the temperature dependence of $1/T_{1}T$ is interpreted as thermal excitation originating from specific band structure 
owing to the disappearance of the hole Fermi surfaces by electron doping~\cite{Ike2}. 
In BKFA, Fermi surfaces change with increasing $x$: hole Fermi surfaces at around the $\Gamma$ point in the BZ becomes larger, 
and electron Fermi surfaces at around the X point shrinks significantly or disappears. 
This Fermi surface change is reported by recent ARPES measurements~\cite{Nak1}, 
and it worsen the nesting condition in BKFA. 
Since the parent compound BaFe$_{2}$As$_{2}$ is a compensated metal, it is natural to consider that the gap-like behavior in this system 
has the same origin explained for electron-doped system Ba(Fe$_{1-x}$Co$_{x}$)$_{2}$As$_{2}$. 
However, the detailed change of band structure in hole-doped system is somewhat different from that in electron-doped system~\cite{not1}. 
In Ba(Fe$_{1-x}$Co$_{x}$)$_{2}$As$_{2}$, hole Fermi surfaces disappear with electron doping, 
and their density of states near Fermi level also decrease with further doping. 
In contrast, in BKFA electron Fermi surfaces do not completely disappear with hole doping, 
but their density of states near Fermi level still exist with further doping until $x$ =1. 
This is confirmed by ARPES measurements~\cite{Sat1,Nak1,Nak2,Yos1,Yos2,Mal1} and also by band structure calculation~\cite{Ike1,Lee1}. 
The recovery of increase of the $1/T_{1}T$ at low temperatures in $x$ = 1 is also indirect evidence for the specific band structure of hole-doped system.

\begin{figure}
\includegraphics[width=8cm]{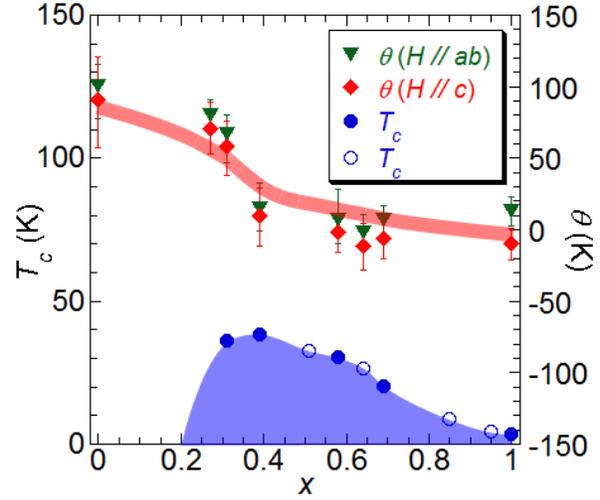}
\caption{
(color online) Phase diagram of BKFA and the $x$ dependence of 
obtained Curie-Weiss temperature $\theta_{\rm CW}$ from the fitting described in the text. 
Red bold line is guide to the eye. 
}
\label{f5}
\end{figure}

The AF spin fluctuation is strong near optimally doped region ($x$ = 0.3-0.4).
This is confirmed by our previous NMR measurements using polycrystalline samples~\cite{Fuk2} and also by neutron diffraction measurements~\cite{Cas1}.
The AF spin fluctuation also exists in heavily overdoped region for $x$ = 0.94 and 1.
Because $T$ independent dynamical susceptibility becomes larger with increasing $x$,
relative 1/$T_{1}T$ for $x$ = 0.94 and 1 is larger than that for $x$ = 0.3-0.4.
From the temperature dependence of $1/T_{1}T$, we can discuss the $\vec{q}$-dependent spin fluctuation of system. 
\begin{equation}\label{T1T}
\frac{1}{T_{1}T} \propto \sum_{\vec{q}}|A_{\vec{q}}|^{2}\frac{\chi_{\bot}^{\prime\prime}(\vec{q},\omega_{0})}{\omega_{0}}
\end{equation} 
Here, $A_{\vec{q}}$ and $\chi_{\bot}^{\prime\prime}(\vec{q},\omega)$ are $\vec{q}$-dependent hyperfine coupling constant and 
perpendicular component against the quantization axis of imaginary part of dynamical susceptibility, respectively. 
Based on the self consistent renormalization theory (SCR) assuming two dimensional AF fluctuation, 
this general expression for $1/T_{1}T$ can be written as follows~\cite{Mor1}. 
\begin{equation}\label{CW}
\frac{1}{T_{1}T} = \chi_{0} + \frac{C}{T-\theta_{\rm CW}}
\end{equation} 
Here, $\chi_{0}$, $C$, and $\theta_{\rm CW}$ are constant term of dynamical susceptibility, Curie constant, and Curie-Weiss temperature, respectively.
$\chi_{0}$ includes all of the temperature independence terms of dynamical susceptibility.
The Curie-Weiss temperature in this definition corresponds to the distance from the AF instability point. 
The value of $\chi_{0}$ strongly influences evaluation of $\theta_{\rm CW}$.
In the present case, we determined $\chi_{0}$ as a product of $1/T_{1}T$ at 300~K and coefficient $\varepsilon(x)$.
$\varepsilon(x)$ reflects convergence of $T_{1}T$ at high temperatures:
typical values of $\varepsilon(x)$ are 0.8 and 0.55 for $x$ = 0 and 1, respectively.
In order to discuss the $x$ dependence of $\theta_{\rm CW}$, we should perform the fitting for the same range of temperature. 
However, as discussed above, gap-like behavior was observed at around 50-100~K for $x$ = 0.58, 0.64, and 0.69. 
Therefore, it is adequate to discuss spin fluctuation in this system using higher temperature data. 
We performed the fitting using the data above 200~K. 
In Fig.~\ref{f5}, we summarized $x$ dependence of $\theta_{\rm CW}$ obtained from the fitting and that of $T_{\rm c}$. 
Because we adopted the eq.~(\ref{CW}) at high temperatures and 
the absolute value of $\theta_{\rm CW}$ has ambiguity due to experimental error of $T_{1}$, 
the relative values and overall feature of the $x$ dependence of $\theta_{\rm CW}$ are important to discuss the spin fluctuation in this system. 
The $x$ dependence of $\theta_{\rm CW}$ indicates that the AF fluctuation becomes weaker with increasing $x$ and that 
the spin fluctuation remains even at the end composition $x$ = 1. 
The tendency that the AF spin fluctuation becomes weaker with increasing $x$ is attributable to 
the poorer nesting condition of Fermi surface with increasing $x$. 
This is consistent with recent reports on inelastic neutron scattering of BKFA~\cite{Lee1,Cas1}. 
The $\theta_{\rm CW}$ obtained at higher temperatures corresponds to the potential AF fluctuation
at high energy excitation, which is responsible for superconductivity, compared with that obtained at lower temperatures. 
These results show that there is a correlation between attracting interaction of superconductivity and AF spin fluctuation to some extent.

\begin{figure}
\includegraphics[width=8cm]{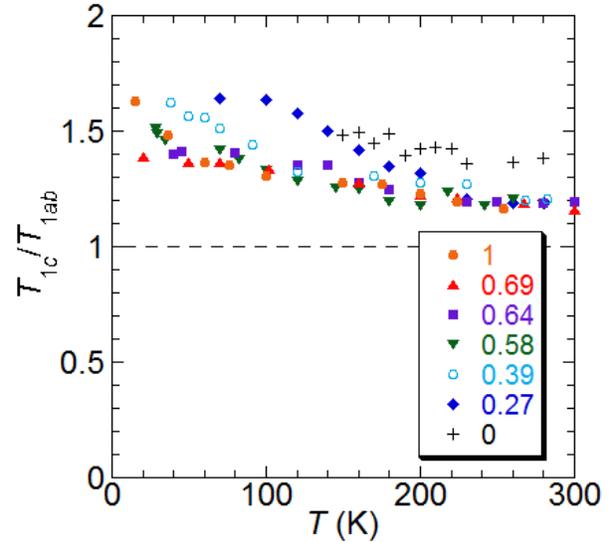}
\caption{
(color online) Temperature dependence of the ratio of $1/T_{1ab}$ to $1/T_{1c}$. 
}
\label{f6}
\end{figure}

Although overall feature of temperature dependence of $1/T_{1}$ is quite similar 
between $1/T_{1}$ with magnetic field parallel to $ab$ plane and that parallel to the $c$ axis. 
However, $1/T_{1ab}(T)$ is always greater than $1/T_{1c}(T)$ at every temperature for all $x$'s. 
In order to discuss this tendency which is related with the dynamical spin susceptibility $\chi(\vec{q},\omega)$, 
we adopt the same procedure described in ref.~\ref{SKita}. 
As described in eq.~(\ref{T1T}), $1/T_{1}T$ is related with the $\vec{q}$-dependent hyperfine coupling constant and dynamical susceptibility. 
This equation can be derived by using fluctuation-dissipation theorem. 
The combination of the hyperfine coupling constant and dynamical susceptibility is intrinsically brought out from hyperfine field at As site. 
By using hyperfine field, $1/T_{1}$ can be rewritten as follows:\\
\begin{eqnarray}\label{T1hyp}
\left( \frac{1}{T_{1}}\right) _{z} &=& 
\frac{(\mu_{0}\gamma_{\rm N})^{2}}{2} \int_{-\infty}^{+\infty}dt \left( \langle H_{{\rm hf},x}(t), H_{{\rm hf},x}(0)\rangle \right. \nonumber\\
 && \qquad\qquad 
\left. +\langle H_{{\rm hf},y}(t), H_{{\rm hf},y}(0)\rangle\right) e^{i\omega_{0}t} \\
 &=& (\mu_{0}\gamma_{\rm N})^{2}\left( |H_{{\rm hf},x}(\omega_{0})|^{2}+|H_{{\rm hf},y}(\omega_{0})|^{2} \right) .
\end{eqnarray} 
Here, direction $z$ corresponds to the direction of external field. 
$H_{{\rm hf},x}$ and $H_{{\rm hf},y}$ are hyperfine fields perpendicular to the $z$ direction. 
We assume that the hyperfine field $\vec{H}_{\rm hf}^{\rm As}$ at As site is brought out from 
spin $\vec{S}$ originating from the nearest neighboring four Fe sites through the hyperfine interaction tensor $\tilde{A}$. 
\begin{eqnarray}\label{Hyp}
\vec{H}_{\rm hf}^{\rm As} = \sum_{i=1}^{4}\tilde{B}_{i}\vec{S}_{i} = \tilde{A}\vec{S}\\ 
\tilde{A} = \left(
            \begin{array}{@{\,}ccc@{\,}}
             A_{a} & D     & B_{1} \\
             D     & A_{b} & B_{2} \\
             B_{1} & B_{2} & A_{c} 
            \end{array}
            \right) 
\end{eqnarray} 
Here,  $\vec{S}_{i}$ and $\tilde{B}_{i}$ are spin at the $i$-th Fe site and 
hyperfine coupling interaction tensor between the $i$-th Fe site and As site, respectively. 
$A_{i}$ ($i = a, b, c$), $B_{i}$ ($i = 1, 2$), and $D$ are diagonal component of $\tilde{A}$ along the $i$ direction, 
off diagonal component related with wave vector $\vec{q} = (\pi,0)$ or $(0,\pi)$, 
and off diagonal component related with wave vector $\vec{q} = (\pi,\pi)$, respectively~\cite{not2}. 
Note that the wave vector $\vec{q} = (\pi,0)$ or $(0,\pi)$ used here corresponds to the wave vector $\vec{q} = (\pi,\pi)$ 
which was experimentally found in inelastic neutron diffraction study~\cite{Lee1}. 
By using these equations and assuming that diagonal components of the tensor and spin are equivalent within $ab$ plane 
($A_{ab}=A_{a}=A_{b}$, $S_{ab}=S_{a}=S_{b}$), we obtain relation between the ratio $R\equiv T_{1c}/T_{1ab}$ and spin $\vec{S}$. 
\begin{subnumcases}
{\textstyle R =}
0.5+0.5\left( \frac{A_{c}S_{c}}{A_{ab}S_{ab}}\right) ^{2} : {\rm no\; correlation} & \label{T1RA} \\
0.5+ \left( \frac{S_{ab}}{S_{c}}\right) ^{2} \qquad : (\pi,0)\; {\rm or}\; (0,\pi) & \label{T1RB}  \\
0.5 \qquad\qquad\qquad : (\pi,\pi) & 
\end{subnumcases}
As discussed in the subsection of Knight shifts, there is a tendency that $A_{ab}$ and $K_{ab}(T)$ are 
greater than $A_{c}$ and $K_{c}(T)$ for entire $x$'s, respectively. 
Hence, we may assume that $S_{ab}$ is also greater than $S_{c}$. 
This leads that the value of eq.~(\ref{T1RA}) is less than 1 and that the value of eq.~(\ref{T1RB}) is greater than 1.5. 
In Fig.~\ref{f6}, we plotted temperature dependence of the ratio $R$. 
$R$ is greater than 1 for all $x$'s within error bar.
This result indicates that spin fluctuation with the wave vector $\vec{q} = (\pi,0)$ or $(0,\pi)$ or wave vector near such wave vector, 
which corresponds to the wave vector $\vec{q} = (\pi,\pi)$ with the notation in the BZ for tetragonal {\it I4/mmm} structure, 
is dominant in entire $x$ composition of BKFA. 
Note that the current result anisotropy of 1/$T_{1}$ for optimally doped composition $x = 0.39$ is consistent with the report for $x$ = 0.32 by Li {\it et al.}~\cite{Li1}. 
Moreover, conclusion obtained in this subsection is consistent with the recent inelastic neutron diffraction studies~\cite{Lee1, Cas1}.

\subsection{$1/T_{1}$ in superconducting state}

\begin{figure}
\includegraphics[width=8cm]{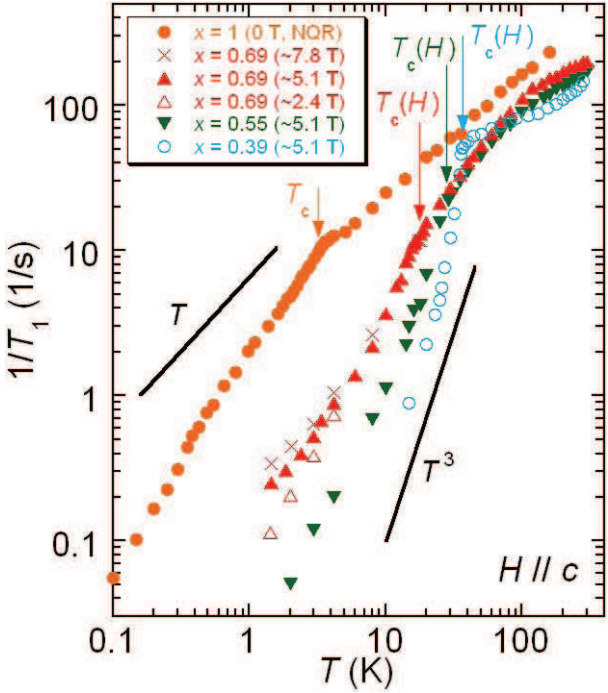}
\caption{
(color online) Temperature dependence of spin-lattice relaxation rate $1/T_{1}$ of $^{75}$As of BKFA. 
}
\label{f7}
\end{figure}

In Fig.~\ref{f7}, we show $T$ dependence of 1/$T_{1}$ for $x$ = 0.39, 0.58, 0.69 and 1.
Each arrow indicates $T_{\rm c}$ under external field.
1/$T_{1}$ has no coherence peak just below $T_{\rm c}$ and decreases rapidly.
$T$ dependence of 1/$T_{1}$ from $T_{\rm c}$ to 0.8$T_{\rm c}$ is approximately proportional to $T^{7}$ for $x$ = 0.39, 
$T^{3.2}$ for $x$ = 0.58, $T^{2.2}$ for $x$ = 0.69.
The power-law behavior was also observed at lower temperature.
The exponent $\alpha$ of power-law $T^{\alpha}$ decreases with increasing $x$.
1/$T_{1}$ for $x$ = 0.39 could not be obtained at lower temperatures,
because the signal intensity dramatically decreases and $T_{1}$ becomes longer in SC state.

1/$T_{1}$ for $x$ = 0.69 under 5.1~T varies proportional to $T$ at lowest temperatures.
With increasing external field, $T$ linear region expands.
And $T$ linear behavior disappeared under 2.4~T.
1/$T_{1}$ for $x$ = 0.69 is consistent with the report of Zhang {\it et al.}~\cite{Zha1}, although steplike feature just below $T_{\rm c}$ was not observed.

In order to understand the change of gap structure, we analyze 1/$T_{1}$ in SC state with two gap model.
For a simplicity we assume that both gaps have same symmetry such as full gapped $s_{\pm}$ or line node.
Here, we assume line node model for $s_{\pm}$-wave with line node or $d$-wave symmetries.
The $1/T_{1}$ in SC state is expressed as,
$$ \frac{1}{T_{1}} \propto \sum_{i=1,2} n^{2}_{i} \int_0^{\infty} \{ N^{i}_S(E)^{2}+M^{i}_S(E)^{2} \} f(E) \{ 1-f(E) \} dE ,$$
where $N^{i}_S(E)^{2}$, $M^{i}_S(E)^{2}$, $f(E)$ are the density of states (DOS), the anomalous DOS arising from the coherence effect of Cooper paris, and the Fermi distribution functions, respectively.
$n_{i}$ presents the fraction of DOS of the $i$-th gap and $n_{1}+n_{2} = 1$.
This is the same procedure utilized in Refs. \ref{Yam20} and \ref{Mata1}.

\begin{table}[h]
\caption{\label{t1}Fitted parameters}
\begin{center}
\begin{tabular}{cccccc}
\hline
Gap type&$x$&$\scriptstyle 2\Delta_{1}(0)/T_{\rm c}$&$\scriptstyle 2\Delta_{2}(0)/T_{\rm c}$&$n_{1}$&$\scriptstyle \delta_{i}/\Delta_{i}$\\
\hline
$s_{\pm}$&0.39&9.6&2.5&0.76&0.1\\
$s_{\pm}$&0.58&5.8&0.92&0.75&0.1\\
$s_{\pm}$&0.69&4.8&0.62&0.65&0.1\\
$s_{\pm}$&1.0&4.0&0.54&0.51&0.1\\
$d$-wave&0.39&14.2&4.4&0.80&-\\
$d$-wave&0.58&8.1&0.91&0.75&-\\
$d$-wave&0.69&5.6&0.66&0.64&-\\
$d$-wave&1.0&4.2&0.48&0.50&-\\
\hline
\end{tabular}
\end{center}
\end{table}

\begin{figure}
\includegraphics[width=8cm]{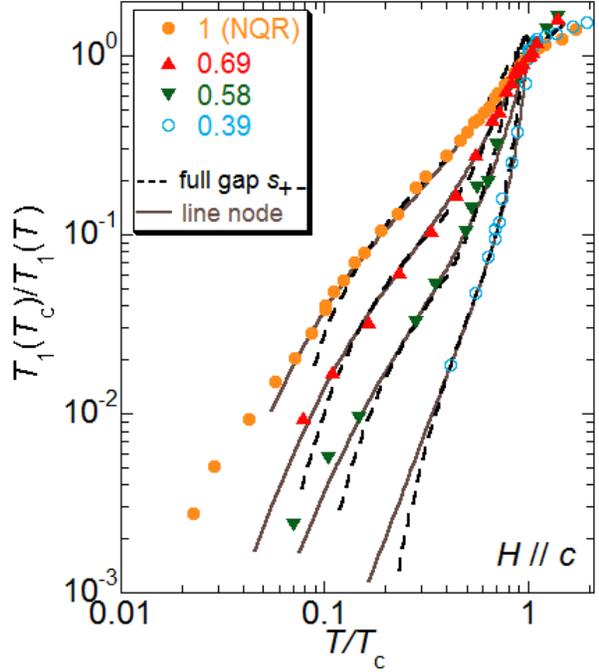}
\caption{
(color online) Normalized temperature dependence of normalized spin-lattice relaxation rate $1/T_{1}$ at $T_{\rm c}$ of 
$^{75}$As of BKFA.
Black dashed lines and brown solid lines are calculated 1/$T_{1}$ of fully gapped $s_{\pm}$ model and line node model, respectively.
}
\label{f8}
\end{figure}

In Fig.~\ref{f8}, we show experimentally obtained 1/$T_{1}$ and calculated 1/$T_{1}$.
Each 1/$T_{1}$ is normalized by the value at $T_{\rm c}$ in order to clarify $x$ dependence of 1/$T_{1}$.
The experimental data for $x=1$ was quoted from Ref.\cite{Fuk1} using poly crystals, 
because we could not perform NMR/NQR measurements using single crystals owing to the drastic reduction of NMR signals in SC state. 
Black dashed lines are obtained with fully gapped $s_{\pm}$ model and brown solid lines with line node model.
The best fitted parameters are listed in Table.~\ref{t1}.
Although we neglected the impurity effect,
the obtained gap parameters for $x$ = 0.39 are consistent with the previous report 
on self-flux-grown single crystal for $x = 0.32$~\cite{Li1}.
Both models can well reproduce the observed 1/$T_{1}$ behaviors in SC state, although the fitting at low temperatures is worse than that at higher temperatures.

\begin{figure}
\includegraphics[width=8cm]{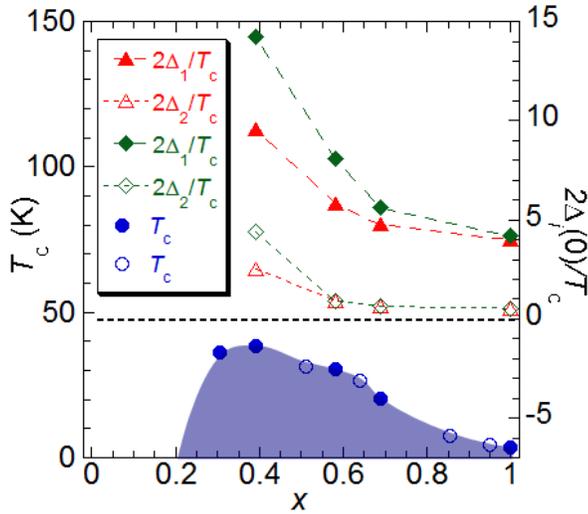}
\caption{
(color online) The $x$ dependence of analyzed superconducting gap parameter $\Delta$ divided by $T_{\rm c}$ 
of Ba$_{1-x}$K$_{x}$Fe$_{2}$As$_{2}$. 
Solid triangles, open triangles, solid diamonds, and open diamonds denote 
$2\Delta_{1}/T_{\rm c}$ for full gap model, $2\Delta_{2}/T_{\rm c}$ for full gap model, 
$2\Delta_{1}/T_{\rm c}$ for nodal-line gap model, and $2\Delta_{2}/T_{\rm c}$ for nodal-line gap model, respectively.
Black dashed line shows zero of $2\Delta_{i}/T_{\rm c}$.
}
\label{f9}
\end{figure}

In Fig.~\ref{f9}, we show the $x$ dependence of analyzed superconducting gap parameter $\Delta$ divided by $T_{\rm c}$ of BKFA. 
The $x$ dependence of $2\Delta_{i}/T_{\rm c}$ for both $i$'s shows monotonous decrease with increasing $x$.
The smaller gap $2\Delta_{2}/T_{\rm c}$ rapidly decreases from $x$ = 0.39 to 0.58.
This probably originates from the poorer nesting condition which arises from the change of band structure by hole doping.
Smaller gap values above $x$ = 0.58 were strongly suppressed.
It corresponds to concentrations where the gap behavior is observed.
The $x$ dependence of $2\Delta_{2}/T_{\rm c}$ is consistent with the recent results of ARPES of BKFA~\cite{Mal1}.
The larger gap $2\Delta_{1}/T_{\rm c}$ is related with the inclination of 1/$T_{1}$ just below $T_{\rm c}$. 
Compared with the $x$ dependence of $2\Delta_{2}/T_{\rm c}$, 
the $x$ dependence of $2\Delta_{1}/T_{\rm c}$ is more gradual and the magnitude decreases continuously.
Therefore, we may conclude that there is no SC symmetry change in BKFA and 
that the nodal-line SC gap structure realized in $x$ = 1 should be explained 
with the same SC symmetry as that realized in optimally doped region $x \sim$ 0.4.
We may speculate that nodal-line structure emerges from $x \sim$ 0.7 and 
that it develops gradually without the change of SC gap symmetry in this system.
However, we cannot completely deny the possibility of the SC gap symmetry change at around $x \sim$ 0.7 
since the numerical calculation suggests that the SC condensation energy for $d$-wave nodal-line gap symmetry 
is slightly lower than that for fully-gapped $s$-wave gap symmetry and 
that $T_{\rm c}$ and the SC gap $2\Delta_{i}/T_{\rm c}$ change continuously with doping in such case~\cite{Suz2,Tho1}.

The gap symmetry cannot be determined solely with NMR measurements, which can be clearly understood from the calculated results. 
To determine gap symmetry, not only NMR/NQR experiments but also other experiments such as specific heat, 
low temperature ARPES, are necessary as suggested in our previous report~\cite{Fuk1}.
Recent ARPES experiment of KFA observed vertical node on zone central hole Fermi surfaces.~\cite{Oka1}.
Since BKFA has complicated band and gap structures,
theoretical interpretations which can explain these results without inconsistency are required.

Finally, we discussed the origin of $T$ linear behavior for $x$ = 0.69 with increasing magnetic field at lowest measurement temperature. 
According to the obtained gap values, ratio of $\Delta_{1}$ to $\Delta_{2}$ is nearly 8.
When we assume the upper critical field of SC ($H_{\rm c2}$) is proportional to $T_{\rm c}$($\sim 20$~K),
$H_{\rm c2}$ of $\Delta_{1}$ would be 20~T or more than this value. 
Thus $H_{\rm c2}$ of $\Delta_{2}$ becomes equal to or more than approximately 2.5~T.
By this estimation it is natural to consider that the smaller gap is collapsed by the external magnetic field 
and that $T$ linear behavior under higher magnetic fields at lower temperatures in SC state was observed.

\section{Conclusion}

In summary, we performed NMR/NQR measurement for BKFA.
The $x$ dependence of the experimental $\nu_{\rm Q}$ is linear increase with increasing concentration $x$.
This tendency is consistent with obtained $\nu_{\rm Q}$ by the electronic structure calculation using WINE2K code. 

$T$ dependence of NMR spectra for $x$ = 0.27 revealed that the paramagnetic region and antiferromanetically ordered region are phase-separated.
Knight shift in normal state exhibits weak temperature dependence for all $x$.
With increasing $x$, Knight shift tends to increase due to the increase of DOS at the Fermi level with hole doping.
Knight shift in SC state exhibits decease below $T_{\rm c}(H)$ for $x$ = 0.39, 0.64, 0.69 and 
the spin part of SC paring symmetry is also singlet in the heavily hole-doped region.

$T$ dependence of 1/$T_{1}T$ indicates that AF spin fluctuation exist for all $x$.
The gap-like temperature dependence for $x$ = 0.58, 0.64, 0.69 is also observed.
The gap-like behavior has probably the same origin explained for electron-doped system.
In order to estimate the strength of spin fluctuation, we performed the fitting from 200 to 300~K using 2D SCR theory.
The $x$ dependence of $\theta_{\rm CW}$ indicates that the spin fluctuation 
becomes weaker with increasing $x$ and that the spine fluctuation remains even at $x$ = 1.
The spin fluctuation tends to become weaker with increasing $x$. 
This tendency also relates with the $x$ dependence of $T_{\rm c}$.
The ratio experimental $(1/T_{1})_{ab}$ to $(1/T_{1})_{c}$ is greater than 1 in all $T$ for all $x$.
This indicates that the wave vector $\vec{q} = (\pi,\pi)$ with the notation in the BZ for tetragonal structure is dominant in entire $x$.
Therefore, we may conclude that there is correlation between attracting interaction of superconductivity and AF spin fluctuation to some extent. 

$T$ dependence of 1/$T_{1}$ in SC state has no coherence peak just below $T_{\rm c}$ and decreases rapidly.
In order to understand the change of gap structure,
we calculate 1/$T_{1}$ in SC state with two gap model with $s_{\pm}$ and nodal-line symmetry cases.
Both symmetry model can well explain in this system, although SC symmetry could not be determined by only NMR experiments.
Both obtained larger gap $\Delta_{1}/k_{\rm B}T_{\rm c}$ and 
smaller gap $\Delta_{2}/k_{\rm B}T_{\rm c}$ rapidly decrease from $x$ = 0.39 to 0.58.
This probably originates from the poor nesting condition arising from the change of band structure.
Furthermore, $\Delta_{2}/k_{\rm B}T_{\rm c}$ is strongly suppressed above $x$ = 0.58.
The $x$ dependence of $\Delta_{1}/k_{\rm B}T_{\rm c}$ is decreases continuously with increasing $x$.
Therefore, it can be concluded that there is basically no SC symmetry change in BKFA.
The nodal-line SC gap structure may be explained with the same SC symmetry in optimum doped region.
According to the analysis for $1/T_{1}$ for $x=0.69$ under magnetic fields, we also conclude that 
the smaller gap is collapsed by the external field at lower temperatures.

\section*{Acknowledgments}

The authors thank K. Ohishi, Y. Ishii, I. Watanabe, K. Okazaki, W. Malaeb, Y. Oota, S. Shin, S. Kittaka, Y. Aoki, and T. Sakakibara for 
fruitful discussion and permitting the authors to refer their unpublished data. 
This work is supported by Grants-in-Aid for Scientific Research (Nos. 21540351 \& 22684016) 
from the Ministry of Education, Culture, Sports, Science and Technology (MEXT) 
and the Japan Society for the Promotion of Science (JSPS), and Innovative Areas ``Heavy Electrons" (Nos. 20102005 \& 21102505) from MEXT, 
Global COE and AGGST financial support program from Chiba University. 
This work at Ames Laboratory was supported by the Division of Material Sciences and Engineering,
Office of Basic Energy Sciences, U.S. Department of Energy.
Ames Laboratory is operated for US Department of Energy by Iowa State University under contract No. DE-AC02-07CH11358.

\newpage

\begin{thebibliography}{99}


\bibitem{Kam1} Y. Kamihara, T. Watanabe, M. Hirano, and H. Hosono: J. Am. Chem. Soc. {\bf 130} (2008) 3296. 

\bibitem{Ren1} Z. A. Ren, J. Yang, W. Lu, W. Yi, G. C. Che, X. L. Dong, L. L. Sun and Z. X. Zhao: DOI 10.1179/143307508X333.686 (2008).

\bibitem{Miy1} K. Miyazawa, K. Kihou, P. M. Shirage, C. H. Lee, H. Kito, H. Eisaki, and A. Iyo, J. Phys. Soc. Jpn. {\bf 78} (2009) 03712. 

\bibitem{Rot1}\label{Rott} M. Rotter, M. Pangerl, M. Tegel: and D. Johrendt: Angew. Chem. Int. Ed. {\bf 47} (2008) 7949. 

\bibitem{Sef1} A. S. Sefat, R. Jin, M. A. McGuire, B. C. Sales, D. J. Singh, and D. Mandrus: Phys. Rev. Lett. {\bf 101} (2008) 117004. 


\bibitem{Jia1} S.Jiang, H. Xing, G. Xuan,, C. Wang, Z. Ren. C. Feng, J. Dai, Z. Xu, and G. Cao: J. Phys.: Condens. Matter {\bf 21} (2009) 382203. 

\bibitem{Kas1} S. Kasahara, T. Shibauchi, K. Hashimoto, K. Ikada, S. Tonegawa, R. Okazaki, H. Shishido, H. Ikeda, H. Takeya, 
K. Hirata, T. Terashima, and Y. Matsuda: Phys. Rev. B {\bf 81} (2010) 184519. 

\bibitem{Ali1} P. L. Alireza, Y T Chris Ko, J. Gillett, C. M. Petrone, J. M. Cole, G. G Lonzarich, and S. E. Sebastian: J. Phys.: Condens. Matter. Jpn. {\bf 21} (2009) 012208. 

\bibitem{Yam1} T. Yamazaki, N. Takeshita, R. Kobayashi, H. Fukazawa, Y. Kohori, and K. Kihou: Phys. Rev. B {\bf 81} (2010) 224511. 

\bibitem{Rei1} J.-Ph. Reid, M. A. Tanatar, X. G. Luo, H. Shakeripour, N. Doiron-Leyraud, N. Ni, S. L. Bud'ko, P. C. Canfield, R. Prozorov, and L. Taillefer: Phys. Rev. B {\bf 82} (2010) 64501.

\bibitem{Yam10} M. Yamashita, Y. Senshu, T. Shibauchi, S. Kasahara, K. Hashimoto, D. Watanabe, H. Ikeda, T. Terashima, I. Vekhter, A. B. Vorontsov, and Y. Matsuda: Phys. Rev. B {\bf 84} (2001) 060507(R).

\bibitem{Suz1} K. Suzuki, H. Usui, and K. Kuroki: J. Phys. Soc. Jpn. {\bf 80} (2011) 013710.

\bibitem{Has2} K. Hashimoto, T. Shibauchi, S. Kasahara, K. Ikada, S. Tonegawa, T. Kato, R. Okazaki, C. J. van der Beek, M. Konczykowski, H. Takeya, K. Hirata, T. Terashima, and Y. Matsuda: Phys. Rev. Lett. {\bf 102} (2009) 207001.

\bibitem{Kha1} R. Khasanov, D. V. Evtushinsky, A. Amato, H. -H. Klauss, H. Luetkens, Ch. Niedermayer, B. B\"{u}chner, G.L. Sun, C. T. Lin, J.T. Park, D. S. Inosov, V. Hinkov: Phys. Rev. Lett. {\bf 102} (2009) 187005.

\bibitem{Din1} H. Ding, P. Richard, K. Nakayama, K. Sugawara, T. Arakane, Y. Sekiba, A. Takayama, 
S. Souma, T. Sato, T. Takahashi, Z. Wang, X. Dai, Z. Fang, G. F. Chen, J. L. Luo and N. L. Wang: 
Euro. Phys. Lett. {\bf 83} (2008) 47001.



\bibitem{Maz1} I. I. Mazin, D. J. Singh, M. D. Johannes, and M. H. Du: Phys. Rev. Lett. {\bf 101} (2008) 057003.

\bibitem{Ike1} H. Ikeda: J. Phys. Soc. Jpn. {\bf 77} (2008) 123707. 

\bibitem{Nag1} Y. Nagai, N. Hayashi, N. Nakai, H. Nakamura, M. Okumura, and M. Machida: N. J. Phys {\bf 10} (2008) 103026.


\bibitem{Suz2} K. Suzuki, H. Usui, and K. Kuroki: arXiv:1108.0657v1.


\bibitem{Yam20} \label{Yam20} M. Yashima, H. Nishimura, H. Mukuda, Y. Kitaoka, K. Miyazawa, P. M. Shirage, K. Kiho, H. Kito, H. Eisaki, and A. Iyo:
J. Phys. Soc. Jpn {\bf 78} (2009) 103702.

\bibitem{Mata1} \label{Mata1} K. Matano, Z. A. Ren, X. L. Dong, L. L. Sun, Z. X. Zhao, and G.-q. Zheng: Euro. Phys. Lett. {\bf 87} (2009) 27012.


\bibitem{Ona1} S. Onari and H. Kontani: Phys. Rev. Lett. {\bf 103} (2009) 177011.


\bibitem{San1} K. Sano and Y. Ono: J. Phys. Soc. J {\bf 78} (2009) 124706.


\bibitem{Yan1} Y. Yanagi, Y. Yamakawa, and Y. Ono: Phys. Rev. B {\bf 81} (2010) 054518.

\bibitem{Ona2} S. Onari and H. Kontani: arXiv:1105.6233v1.

\bibitem{Kon10} H. Kontani, T. Saito, and S. Onari: Phys. Rev. B {\bf 84} (2011) 024528.


\bibitem{Fuk1} H. Fukazawa, Y. Yamada, K. Kondo, T. Saito, Y. Kohori, K. Kuga, Y. Matsumoto, S. Nakatsuji, H. Kito, P. M. Shirage, 
K. Kihou, N. Takeshita, 
C. H. Lee, A. Iyo, and H. Eisaki: J. Phys. Soc. Jpn. {\bf 78} (2009) 083712. 

\bibitem{Has1} K. Hashimoto, A. Serafin, S. Tonegawa, R. Katsumata, R. Okazaki, T. Saito, H. Fukazawa, Y. Kohori, K. Kihou, C. H. Lee, A. Iyo, H. Eisaki, H. Ikeda, Y. Matsuda, A. Carrington, and T. Shibauchi: Phys. Rev. B {\bf 82} (2010) 014526. 

\bibitem{Don1} J. K. Dong, S. Y. Zhou, T. Y. Guan, H. Zhang, Y. F. Dai, X. Qiu, X. F. Wang, Y. He, X. H. Chen, and S. Y. Li: 
Phys. Rev. Lett. {\bf 104} (2010) 087005.

\bibitem{Kaw1} H. Kawano-Furukawa, C. J. Bowell, J. S. White, R. W. Heslop, A. S. Cameron, E. M. Forgan, K. Kihou, C. H. Lee, A. Iyo, H. Eisaki, T. Saito, H. Fukazawa, Y. Kohori, R. Cubitt, C. D. Dewhurst, J. L. Gavilano, and M. Zolliker: Phys. Rev. B {\bf 84} (2011) 024507.

\bibitem{KOh1} K. Ohishi, Y. Ishii, H. Fukazawa, T. Saito, I. Watanabe, Y. Kohori, T. Suzuki, K. Kihou, C. H. Lee, K. Miyazawa, H. Kito, A. Iyo, and H. Eisaki: arXiv:1112.6078.

\bibitem{Oka1} K. Okazaki: private communication.

\bibitem{Aok1} Y. Aoki: private communication.





\bibitem{Tho1} R. Thomale, C. Platt, W. Hanke, J. Hu, and B. A. Bernevig: Phys. Rev. Lett. {\bf 107} (2011) 117001.

\bibitem{Lee1} C. H. Lee, K. Kihou, H. Kawano-Furukawa, T. Saito, A. Iyo, H. Eisaki, H. Fukazawa, Y. Kohori, K. Suzuki, 
H. Usui, K. Kuroki, K. Yamada: Phys. Rev. Lett. {\bf 106} (2011) 067003.

\bibitem{Sat1} T. Sato, K. Nakayama, Y. Sekiba, P. Richard, Y.-M. Xu, S. Souma, T. Takahashi, G. F. Chen, J. L. Luo, N. L. Wang, and H. Ding: 
Phys. Rev. Lett. {\bf 103} (2009) 047002.

\bibitem{Ter1} T. Terashima, M. Kimata, N. Kurita, H. Satsukawa, A. Harada, K. Hazama, M. Imai, A. Sato, K. Kihou, 
C. H. Lee, H. Kito, H. Eisaki, A. Iyo, T. Saito, 
H. Fukazawa, Y. Kohori, H. Harima, and S. Uji: J. Phys. Soc. Jpn. {\bf 79} (2010) 053702.





\bibitem{Nak1} K. Nakayama, T. Sato, P. Richard, Y.-M. Xu, T. Kawahara, K. Umezawa, T. Qian, M. Neupane, G. F. Chen, H. Ding, and T. Takahashi:
Phys. Rev. B {\bf 83} (2011) 020501.

\bibitem{Kih1}\label{Kihou} K. Kihou, T. Saito, S. Ishida, M. Nakajima, Y. Tomioka, H. Fukazawa, Y. Kohori, T. Ito, S.-I. Uchida, 
A. Iyo, C. H. Lee, and H. Eisaki: J. Phys. Soc. Jpn. {\bf 79} (2010) 124713. 

\bibitem{Fuk2} H. Fukazawa, T. Yamazaki, K. Kendo, Y. Kohori, N. Takeshita, P. M. Shirage, K. Kihou, K. Miyazawa, H. Kito, H. Eisaki, and A. Iyo: 
J. Phys. Soc. Jpn. {\bf 78} (2009) 033704.


\bibitem{Bas1} T. J. Bastow: J. Phys.: Condens. Matter {\bf 11} (1999) 569.

\bibitem{Bla1} P. Blaha, K. Schwarz, G. K. H. Madsen, D. Kvasnicka, and J. Luitz:
{\it WIEN2K, An Augmented Plane Wave Plus Local OrbitalsProgram for Calculating Crystal Properties}, 
edited by K. Schwarz (Techn. Universit$\ddot{\rm a}$t Wien, Austria, 2001).

\bibitem{Sin1} D. J. Singh: Phys. Rev. B {\bf 79} (2009) 174520. 

\bibitem{Kit1}\label{Kitag} K. Kitagawa, N. Katayama, K. Ohgushi, M. Yoshida, and M. Takigawa: J. Phys. Soc. Jpn. {\bf 77} (2008) 114709.

\bibitem{Fuk3} H. Fukazawa, T. Saito, Y. Yamada, K. Kondo, M. Hirano, Y. Kohori, K. Kuga, A. Sakai, 
Y. Matsumoto, S. Nakatsuji, K. Kihou, A. Iyo, C. H. Lee, and H. Eisaki: J. Phys. Soc. Jpn. {\bf 80} (2011) SA118. 

\bibitem{Wie1} E. Wiesenmayer, H. Luetkens, G. Pascua, R. Khasanov, A. Amato, H. Potts, B. Banusch, H. -H. Klauss, and D. Johrendt:
arXiv:1108.4307.

\bibitem{Nakai1} Y. Nakai, T. Iye, S. Kitagawa, K. Ishida, H. Ikeda, S. Kasahara, H. Shishido, T. Shibauchi, Y. Matsuda, and T. Terashima:
Phys. Rev. Lett. {\bf 105} (2010) 107003.


\bibitem{Nin1} F. Ning, K. Ahilan, T. Imai, A. S. Sefat, R. Jin, M. A. Mcguire, B. C. Sales, and D. Mandrus: 
J. Phys. Soc. Jpn. {\bf 77} (2008) 103705. 

\bibitem{Mata2}\label{Matano} K. Matano, G. L. Sun, D. L. Sun, C. T. Lin, M. Ichioka, G.-q. Zheng: Euro. Phys. Lett. {\bf 87} (2009) 27012.

\bibitem{Sim1} W. W. Simmons, W. J. O'Sullivan, and W. A. Robinson: Rhys. Rev. {\bf 127} (1962) 1168. 


\bibitem{Ike2} H. Ikeda, R. Arita, and J. Kunes, Phys. Rev. B. {\bf 82} (2010) 024508.

\bibitem{not1} Recently new compound K$_{x}$Fe$_{2}$Se$_{2}$ and related compounds, which can be considered to be electron-doped system, 
are found (J.G. Guo, S.F. Jin, G. Wang, S.C. Wang, K.X. Zhu, T.T. Zhou, M. He, and X.L. Chen: Phys. Rev. B {\bf 82} (2010) 180520.). 
However, in this paper, we concentrate on the discussion of hole-doped and electron-doped system which is related to BaFe$_{2}$As$_{2}$ as the parent compound. 


\bibitem{Nak2} 
Y. Sekiba, T. Sato, K. Nakayama, K. Terashima, P. Richard, J. H. Bowen, H. Ding, Y. -M. Xu, L. J. Li, G. H. Cao, Z. -A. Xu, and T. Takahashi: 
New. J. Phys {\bf 11} (2009) 025020.

\bibitem{Yos1} 
W. Malaeb, T. Yoshida, A. Fujimori, M. Kubota, K. Ono, K. Kihou, P. M. Shirage, H. kito, A. Iyo, H. Eisaki, and Y. Kakajima:
J. Phys. Soc. Jpn. {\bf 78} (2009) 123706. 

\bibitem{Yos2} T. Yoshida, I. Nishi, A. Fujimori, M. Yi, R. G. Moore, D.-H. Lu, Z.-X. Shen, K. Kihou, P. M. Shirage, H. Kito, C. H. Lee, 
A. Iyo, H. Eisaki, and H. Harima: arXiv:1007.2698: 
Proceeding of the 9th International Conference on Spectroscopies in Novel Superconductors (SNS2010).



\bibitem{Mal1} W. Malaeb: private communication.




\bibitem{Cas1} J.-P. Castellan, S. Rosenkranz, E. A. Goremychkin, D. Y. Chung, I. S. Todorov, M. G. Kanatzidis, I. Eremin, J. Knolle, 
A. V. Chubukov, S. Maiti, M. R. Norman, F. Weber, H. Claus, T. Guidi, R. I. Bewley, and R. Osborn: Phys. Rev. Lett. {\bf 107} (2011) 177003. 



\bibitem{Mor1} T. Moriya, J. Mag. Mag. Mat. {\bf 100} (1991) 261.  





\bibitem{SKi1} \label{SKita} S. Kitagawa, Y. Nakai, T. Iye, K. Ishida, Y. Kamihara, M. Hirano, and H. Hosono: Phys. Rev. B. {\bf 81} (2010) 212502.


\bibitem{not2} Here, we express wave vector with the notation in the BZ for orthorhombic {\it Fmmm} structure
although actual wave vector should be expressed with that for tetragonal {\it I4/mmm} structure. 
This is because it is complicated to separate the off-diagonal component with wave vector of $(\pi,\pi)$ in the present form 
into those with the wave vector in the form with BZ for the tetragonal {\it I4/mmm} structure

\bibitem{Li1} Z. Li, D. L. Sun, C. T. Lin, Y. H. Su, J. P. Hu, and G. q. Zheng: Phys. Rev. B. {\bf 83} (2011) 140506(R).


\bibitem{Zha1} S. W. Zhang, L. Ma, Y. D. Hou, J. S. Zhang, T. L. Xia, G. F. Chen, J. P. Hu, G. M. Luke, and W. Yu:
Phys. Rev. B. {\bf 81} (2010) 012503.

















\end{thebibliography}
\end{document}